# V viewpoints

Erik Brynjolfsson, Paul Hofmann, and John Jordan

## Economic and Business Dimensions
## Cloud Computing and Electricity: Beyond the Utility Model

*Assessing the strengths, weaknesses, and general applicability of the computing-as-utility business model.*

**B**USINESSES RELY NO less on electricity than on IT. Yet corporations don't need a "Chief Electricity Officer" and a staff of highly trained professionals to manage and integrate electricity into their businesses. Does the historical adoption of electricity offer a useful analogy for today's innovations in cloud computing?

While the utility model offers some insights, we must go beyond this simple analogy to understand cloud computing's real challenges and opportunities. Technical issues of innovation, scale, and geography will confront managers who attempt to take advantage of offsite resources. In addition, business model challenges related to complementarity, interoperability, and security will make it difficult for a stable cloud market to emerge. An overly simplistic reliance on the utility model risks blinding us to the real opportunities and challenges of cloud computing.

### Cloud Computing and the Electricity Model

Definitions for cloud computing vary. From a practitioner standpoint: "Cloud computing is on-demand access to virtualized IT resources that are housed outside of your own data center, shared by others, simple to

> An overly simplistic reliance on the utility model risks blinding us to the real opportunities and challenges of cloud computing.

use, paid for via subscription, and accessed over the Web." From an academic perspective: "Cloud computing refers to both the applications delivered as services over the Internet and the hardware and systems software in the data centers that provide those services. ... The data center hardware and software is what we will call a cloud. When a cloud is made available in a pay-as-you-go manner to the public, we call it a public cloud; the service being sold is utility computing."[1]

Both definitions imply or explicitly use the "utility" model that embeds the logic of water supply, electrical grids, or sewage systems. This model is ubiquitous. While it has important strengths, it also has major weaknesses.

Hardware providers introduced the language of "utility" computing into the market. But perhaps the most rigorous and vigorous assertion of the electricity model comes from Nicholas Carr, an independent blogger in his recent book, *The Big Switch*: "At a purely eco-nomic level, the similarities between electricity and information technology are even more striking. Both are what economists call general-purpose technologies. ... General-purpose technologies, or GPTs, are best thought of not as discrete tools but as platforms on which many different tools, or applications, can be constructed. ... Once it becomes possible to provide the technology centrally, large-scale utility suppliers arise to displace the private providers. It may take decades for companies to abandon their proprietary supply operations and all the investment they represent. But in the end the savings offered by utilities become too compelling to resist, even for the largest enterprises. The grid wins."[4]

### Strengths of the Utility Model

Carr correctly highlights the concept of a general-purpose technology. This class of technology has historically been the greatest driver of productivity growth in modern economies. They not only contribute directly, but also by catalyzing myriad complementary innovations.[3] For electricity, this includes the electric lighting, motors, and machinery. For IT, this includes transaction processing, ERP, online commerce and myriad other applications and even business model innovations.

Some of the economies of scale and cost savings of cloud computing are also akin to those in electricity generation. Through statistical multiplexing, centralized infrastructure can run at higher utilization than many forms of distributed server deployment. One system administrator, for example, can tend over 1,000 servers in a very large data center, while his or her equivalent in a medium-sized data center typically manages approximately 140.[7]

By moving data centers closer to energy production, cloud computing creates additional cost savings. It is far cheaper to move photons over the fiber-optic backbone of the Internet than it is to transmit electrons over our power grid. These savings are captured when data centers are located near low-cost power sources like the hydroelectric dams of the northwest U.S.

Along with its strengths, however, the electric utility analogy also has three technical weaknesses and three business model weaknesses.

### Technical Weaknesses of the Utility Model

*The Pace of Innovation.* The pace of innovation in electricity generation and distribution happens on the scale of decades or centuries.[8] In contrast, Moore's Law is measured in months. In 1976, the basic computational power of a $200 iPod would have cost one billion dollars, while the full set of capabilities would have been impossible to replicate at any price, much less in a shirt pocket. Managing innovative and rapidly changing systems requires the attention of skilled, creative people, even when the innovations are created by others, unlike managing stable technologies.

*The Limits of Scale.* The rapid availability of additional server instances is a central benefit of cloud computing, but it has its limits. In the first place, parallel problems are only a subset of difficult computing tasks: some problems and processes must be attacked with other architectures of processing, memory, and storage, so simply renting more nodes will not help. Secondly, many business applications rely on consistent transactions supported by RDBMS. The CAP Theorem says one cannot have consistency and scalability at the same time. The problem of scalable data storage in the cloud with an API as rich as SQL makes it difficult for high-volume, mission-critical transaction systems to run in cloud environments.

Meanwhile, companies of a certain size can get the best of both worlds by deploying private clouds. Intel, for example, is consolidating its data centers from more than 100 eventually down to about 10. In 2008 the total fell to 75, with cost savings of $95 million. According to Intel's co-CIO Diane Bryant, 85% of Intel's servers support engineering computation, and those servers run at 90%

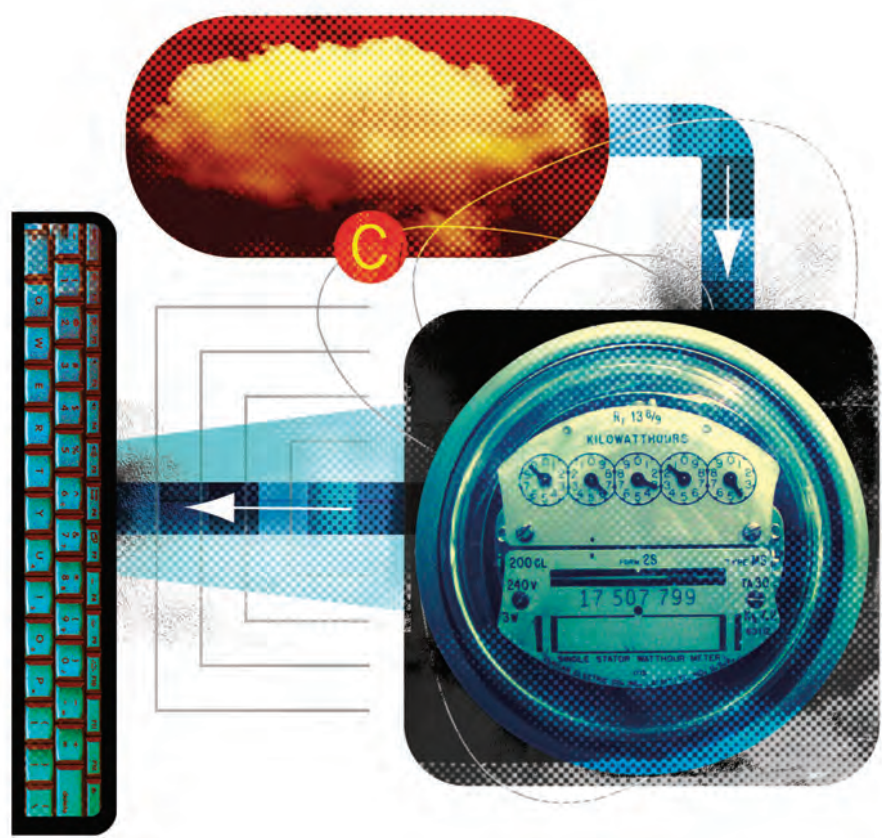

utilization—a combination of strategic importance and operational performance that would negate any arguments for shifting that load to a cloud vendor. Ironically, even as the utility model is being touted for computing, the highly centralized approach is becoming less effective for electricity itself: an emerging distributed power generation system features smaller nodes running micro-hydro, wind, micro-turbines and fuel cells. What's more, many enterprises do in fact generate their own electricity or steam, for the same reasons they will continue to keep certain classes of

ILLUSTRATION BY STUART BRADFORD

IT in house: reliability, strategic advantage, or cost visibility.

*Latency: Distance is Not Dead.* One of the few immutable laws of physics is the speed of light. As a result, latency remains a formidable challenge. In the network realm, the demands for nearly instantaneous execution of machine-to-machine stock trades has led financial services firms to locate their data centers as physically close to stock exchanges as possible. The read/write limits of magnetic disks can only drop so far, but increased speed comes at the cost of capacity: big disks are slow, and fast disks are small. For many classes of applications, performance, convenience, and security considerations will dictate that computing be local. Moving data centers away from their customers may save on electricity costs, but those savings are often outweighed by the costs of latency.

### Beyond Electricity: The Business Model of the Cloud

Important as the technical differences are between electricity and cloud computing, the business model differences are even more profound.

*Complementarities and Co-invention.* Like electricity, IT is a general-purpose technology. This means that critical benefits come from the co-inventions that the basic technology makes possible. It took 30 to 40 years for the full benefits of electricity to redound to America's factories.[5] Initially, assembly lines and production processes were not redesigned to take advantages of electricity: large central steam engines were simply replaced with large electric motors, and then hooked up to the same old crankshafts and cogs. Only with the reinvention of the production process was the potential of electrification realized. Today, electricity has matured to become a relative commodity. In contrast, computing is still in the midst of an explosion of innovation and co-invention.[2] Firms that simply replace corporate resources with cloud computing, while changing nothing else, are doomed to miss the full benefits of the new technology.

The opportunities, and risks, from IT-enabled business model innovation and organizational redesigns are reshaping entire industries.[3] For instance, Apple's transition from a perpetual license model to the pay-per-use iTunes store helped it quadruple revenues in four years. The tight integration between Apple's ERP system and the billing engine handling some 10 million sales per day would have been difficult, if not impossible, in the cloud.

*Lock-in and Interoperability.* Lock-in issues with electricity were addressed long ago by regulation of monopolies, then later by legal separation of generation from transmission and the creation of market structures. Markets work because electrons are fungible. The rotary converter that enabled interconnection of different generating technologies in the 1890s has no analog for the customer of multiple cloud vendors, and won't anytime soon. For enterprise computing to behave like line voltage will require radically different management of data than what is on anyone's technology roadmap.

Perhaps most critically, bits of information are not electrons. Depending on the application, its engineering, and its intended use, cloud offerings will not be interchangeable across cloud providers. Put more simply, the business processes supported by enterprise computing are not motors or light bulbs.

*Security.* The security concerns with cloud computing have no electricity analog. No regulatory or law enforcement body will audit a company's electrons, but processes related to customer data, trade secrets, and classified government information are all subject to stringent requirements and standards of auditability. The typically shared and dynamic resources of cloud computing (including CPU, networking, and so forth) reduce control for the user and pose severe new security issues not encountered by on-premise computing behind firewalls.

### Conclusion

If the utility model were adequate, the challenges to cloud computing could be solved with electricity-like solutions—but they cannot. The reality is that cloud computing cannot achieve the plug-and-play simplicity of electricity, at least, not as long as the pace of innovation, both within cloud computing itself, and in the myriad applications and business models it enables, continues at such a rapid pace. While electric utilities are held up as models of simplicity and stability, even this industry is not immune from the transformative power of IT.[8,9] Innovations like the "smart grid" are triggering fundamental changes at a pace not seen since the early days of electrification.

The real strength of cloud computing is that it is a catalyst for more innovation. In fact, as cloud computing continues to become cheaper and more ubiquitous, the opportunities for combinatorial innovation will only grow. It is true that this inevitably requires more creativity and skill from IT and business executives. In the end, this not something to be avoided. It should be welcomed and embraced.  C

**Erik Brynjolfsson** (erikb@mit.edu) is a professor at the MIT Sloan School and the director of the MIT Center for Digital Business in Cambridge, MA.

**Paul Hofmann** (paul.hofmann@sap.com) is a vice president at SAP Labs in Palo Alto, CA.

**John Jordan** (jmj13@smeal.psu.edu) is a senior lecturer in the Smeal College of Business at Penn State University.

> If the utility model were adequate, the challenges to cloud computing could be solved with electricity-like solutions—but they cannot.